\documentclass[1p,12pt]{elsarticle}

\usepackage{graphicx}
\usepackage{caption}
\usepackage{amssymb}
\usepackage{url}

\makeatletter
\def\ps@pprintTitle{%
 \let\@oddhead\@empty
 \let\@evenhead\@empty
 \def\@oddfoot{\small{doi.org/10.1016/j.sna.2019.111673}\hfill}%
 \let\@evenfoot\@oddfoot}
\makeatother

\begin{document}

\begin{frontmatter}


\title{Reading thin film permanent magnet irregular patterns using magnetoresistive sensors\tnoteref{license}}
\tnotetext[license]{\copyright 2019. This manuscript is made available under the CC-BY-NC-ND 4.0 license: \url{creativecommons.org/licenses/by-nc-nd/4.0/}}



\author[inesc,ist]{Sofia Abrunhosa\corref{cor1}}
\author[inesc]{Karla J. Merazzo}
\author[bogen]{Tiago Costa}
\author[bogen]{Oliver Sandig}
\author[inesc,ist]{Fernando Franco}
\author[inesc,ist]{Susana Cardoso}

\address[inesc]{INESC - Microsistemas e Nanotecnologias, Rua Alves Redol 9, Lisbon 1000-029, Portugal}
\address[ist]{Instituto Superior Técnico - Universidade de Lisboa, Lisbon, Portugal}
\address[bogen]{BOGEN Electronic GmbH, Potsdamer Straße 12-13, 14163 Berlin, Germany}
\cortext[cor1]{Corresponding author. E-mail address: sabrunhosa@inesc-mn.pt}

\begin{abstract}
Magnetic encoders currently provide accurate positioning information by reading periodic patterns with equally spaced structures or bits. As this technology evolves to fulfill industrial demands for cheaper and more accurate systems, variations of this concept with non-periodic and non-predetermined patterns emerge. In this work, we take the first step towards the development of such devices using equally sized bits with different separations. We show magnetic scans of 1000\ $\mu m$ wide and 2000\ \AA \ thick CoCrPt structures with separations between 63 and 3000\ $\mu m$, obtained with tunnel magnetoresisistive sensors. A theoretical model to describe the magnetic field created by these structures is also proposed, which matches the trend of the experimental signals. We also show that there is a spatial resolution loss for reading distances higher than a threshold that depends on the minimum feature size of the sample.
\end{abstract}


\end{frontmatter}


\pagebreak

\section*{Graphical Abstract}

\begin{figure}[h!]
\label{fig:ga}
\centering\includegraphics[width=0.8\linewidth]{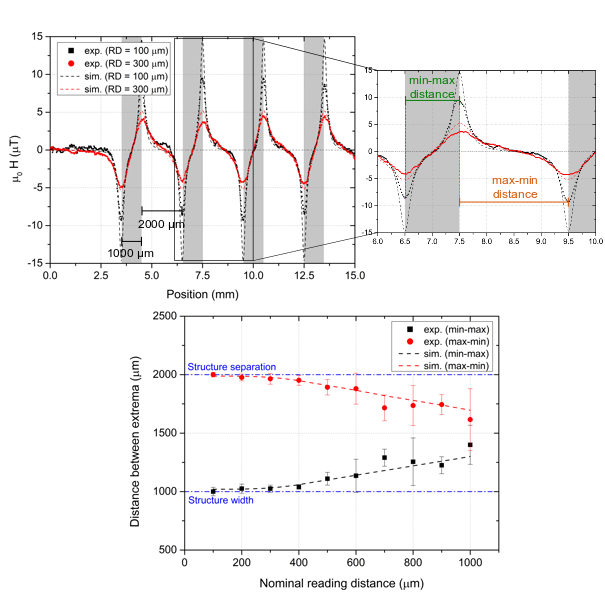}
\caption*{Magnetic scan of a CoCrPt track comprised of 1000 $\mu$m wide structures with separation of 2000 $\mu$m and the evolution of the distance between consecutive extrema with increasing reading distance.}
\end{figure}

\section*{Highlights}
\begin{itemize}
\item Magnetic scales with differently spaced CoCrPt structures have been microfabricated and scanned with magnetoresistive sensors
\item A theoretical model to describe these scales has been developed and is compatible with the experimental results
\item We have observed a decrease in the spatial accuracy of the scans as the reading distance was increased
\end{itemize}

\section{Introduction}
\label{S:1}
Positional encoders are transducers that produce an analog or a digital signal depending on the current position of an object that undergoes a rotational or linear movement. These devices, currently based on optical or magnetic phenomena, are comprised of a track that contains the position information, which is read by the reading head. Commercial optical encoders are generally more accurate, supporting pole-pitches down to 20 $\mu$m \cite{renishaw2019}, although magnetic encoders are usually more robust and capable of operating in dirty and dusty environments \cite{tumanski2016, eitel2014}. Currently, magnetic encoders can be classified as incremental or absolute \cite{miyashita1987}. For incremental systems, the track is made out of a magnetic material coded with periodically alternating magnetic poles - the bits - with equal dimensions \cite{bogen2019, sensitec2019}. These encoders provide information regarding the displacement of an object, i.e. the distance to the starting point of the measurement. On the other hand, absolute measurement systems provide position information that is independent of the starting point and contain more complex encoding patterns with diverse configurations. Magnetic stripes rotated by increasingly higher angles along the length of the track \cite{becker2018} can be used, as well as two adjacent tracks with different pole-pitches, similarly to a Vernier scale \cite{zoschke2018}. Furthermore, 3D printed magnets with varying magnetization along their length have also been shown to be viable for absolute measurement systems \cite{windl2017}.

In this work, we study a variation of the incremental track concept with a different configuration of bits. We use CoCrPt as the track material, largely employed in magnetic tape applications \cite{dee2008, furrer2018}. Merazzo et. al \cite{merazzo2018} have shown previous results with CoCrPt tracks that, instead of containing bits with alternating magnetization, were comprised of 1 mm wide structures magnetized along the same direction and separated by a distance equal to their width. We build upon this design by fabricating CoCrPt samples with 1 mm structures separated by distances comprised in the 63 - 3000 $\mu m$ range. This allows the encoding of more complex information than what is contained in tracks comprised of equally sized and equally spaced bits. These structures are magnetized in-plane along the scanning direction and the vertical component of the resulting magnetic stray field is measured. Moreover, placing two or more of these tracks side-by-side would also open up the construction of an absolute positioning system. At this point, we chose to focus on the incremental version, although the study of the absolute version remains a possibility in the future. 

On the other hand. the reading head of an encoder contains a sensor capable of reading the magnetic field created by the track bits, that is usually a Hall-effect or a magnetoresistive sensor \cite{freitas2007, freitas2016}. Currently, most commercial devices make use of Hall-effect sensors although recent applications have come to include magnetoresistive technology due to their improved spatial accuracy and sensitivity \cite{jander2005, leitao2013, cardoso2014, liu2017}. NDT and magnetic scanning devices using GMR (Giant Magnetoresistance) and TMR (Tunnel Magnetoresistance) have been repeatedly reported in the literature \cite{jin2017, caetano2018}. 

In this work, we focus on characterizing incremental CoCrPt tracks with differently spaced bits using TMR sensors in the reading head. Alongside these sensors, the reading head also includes a set of two permanent magnets (PM) to guarantee that the magnetization of the structures is such that the resulting magnetic field has a component along the sensitive direction of the sensor.

\section{Experimental details}
\label{S:2}
Figure 1 shows the magnetic scanner system utilized, which was initially purposed for non-destructive testing (NDT) \cite{rosado2014, franco2017}. The reading head contains the TMR sensor and the PM and can be moved in the y and z directions to change the position of the scan and the reading distance, respectively. The table itself can be moved in the x direction, to perform each of the scans. Movement in each of the directions is controlled by three stepper motors with a minimum step of 5 $\mu m$, a maximum range of 30 cm in the x and y directions and a maximum height of 10 cm. A detailed view of the reading head is shown in Figure 2a.

\begin{figure}[h!]
\label{fig:1}
\centering\includegraphics[width=0.7\linewidth]{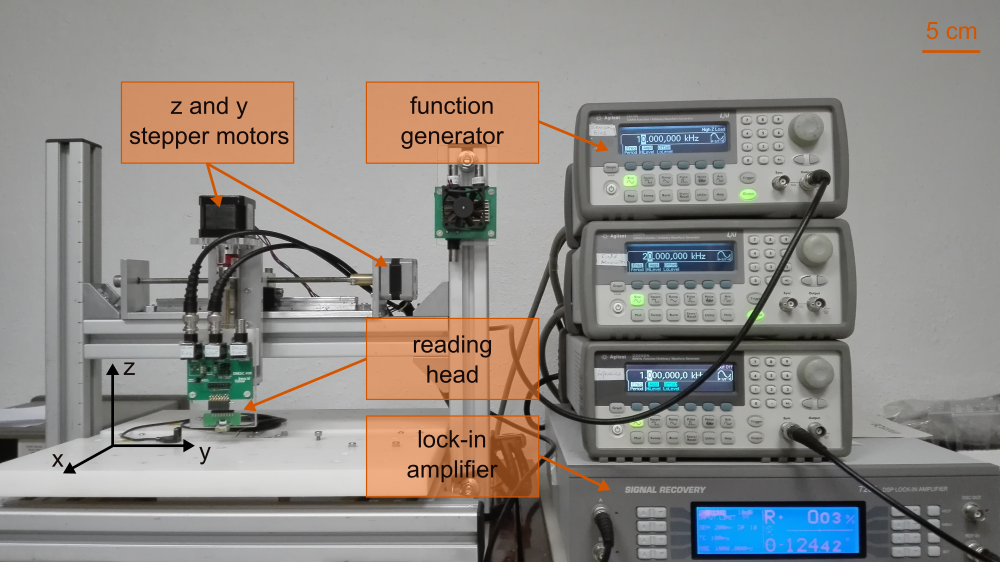}
\caption{Magnetic scanner apparatus. During a scan, the table moves in the x direction. The reading read can be moved in the y and z directions independently to set the scanning position and height.}
\end{figure}

\begin{figure}[h!]
\label{fig:2}
\centering\includegraphics[width=0.8\linewidth]{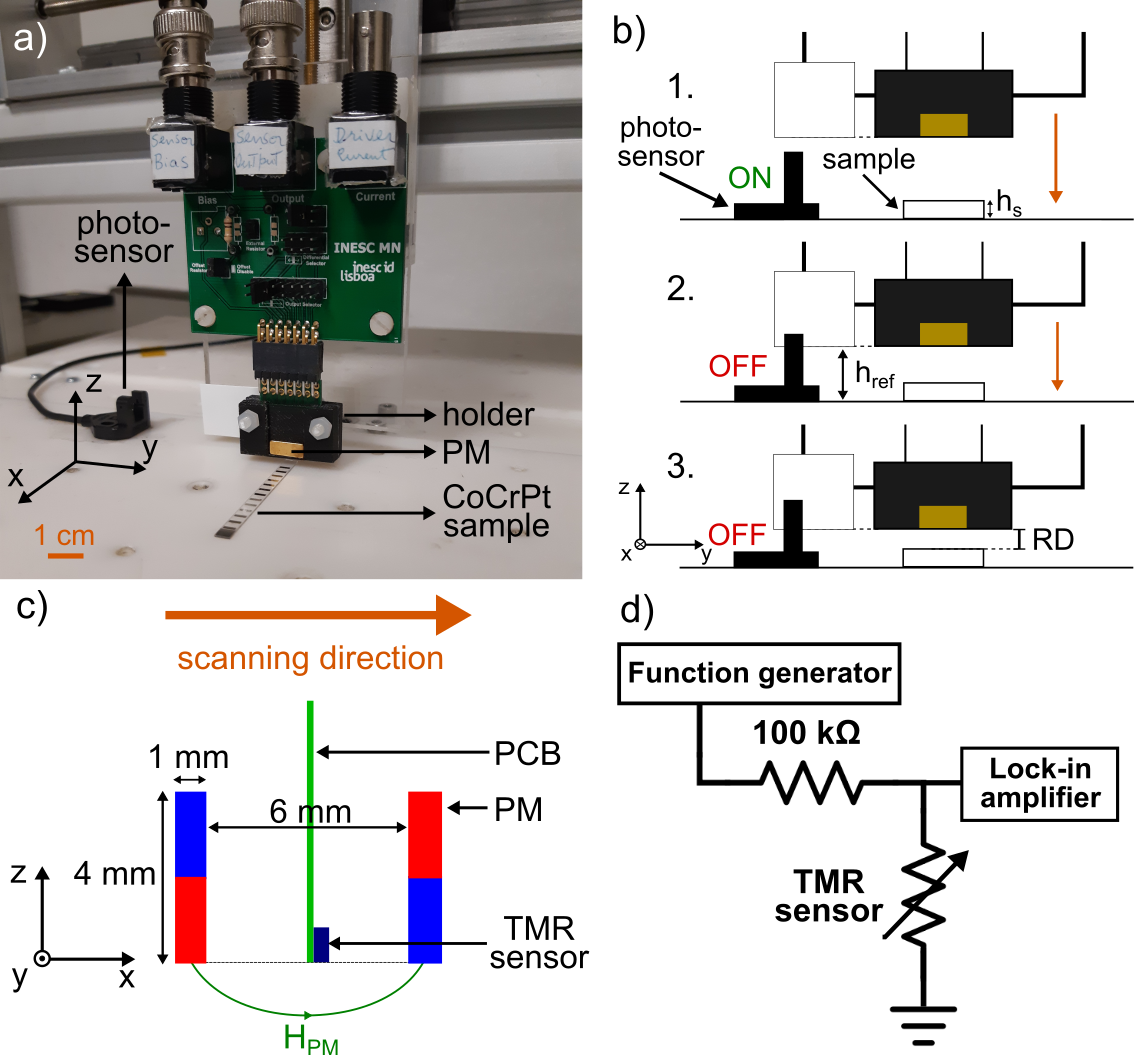}
\caption{a) Detail of the reading head in the magnetic scanner. A second PM has been placed opposite to the visible one and the TMR chip is located in between. b) Schematic representation of the procedure to set the reading distance. c) Schematic representation of the reading head. The permanent magnets are placed in an anti-parallel configuration in regards to their magnetization. d) Biasing circuit for the TMR sensor.}
\end{figure}

The procedure to set the reading distance (distance from the top of the sample to the bottom of the sensor) is represented in Figure 2b. An Omron EE-SX771 containing a photodiode and a photosensor with a precision of 3 $\mu m$ is used, along with a fixed reference placed at sensor level. Firstly, the reading head is positioned so that this reference is directly above the photosensor (step 1) and it is then lowered until the infrared beam is interrupted (step 2). This height is known and is referred to as $h_{ref}$. Knowing the substrate thickness ($h_S$) and the desired reading distance ($RD$), the reading head is then lowered (step 3) by an amount equal to $h_{ref}-h_S-RD$. This method has a precision of 20-30 $\mu m$ and an estimated accuracy of 50-100 $\mu m$. 

As a sensitive element, a TMR sensor was used as described in previous works \cite{paz2016, guo2014}. Each sensor was comprised of 12 elements in series with an individual dimension of $2\times 20\ \mu m^2$. After fabrication, each sensor was placed on a printed circuit board (PCB) and mounted on a 3D printed holder in the yz plane. In order to guarantee that the magnetization of the magnetic structures under the sensor was along the x direction during scanning, two PM with dimensions  $1\times 10\times 4\ mm^3$ and magnetization along their width with a residual magnetism of 1.43 T were also included. These were placed in the same holder using the configuration shown in Figure 2c, which guaranteed that the sensor was able to measure the z component of the magnetic field created by the structures while also minimizing the effect of the PM on the transfer curve. The biasing of the sensor was done using alternate current (AC) with a frequency of 10kHz and an amplitude of 10V provided by an Agilent 33210A function generator, as shown in Figure 2d. For amplification and measuring, a Signal Recovery 7265 lock-in amplifier was used. In this configuration, the sensor had a sensitivity of $2.43\ k\Omega/mT$ ($243\ mV/mT$ with a bias current of 100\ $\mu$A) and the transfer curve shown in Figure 3. An offset field of 0.18 $mT$ and a coercivity of 0.27 $mT$ were observed.

\begin{figure}[h!]
\label{fig:3}
\centering\includegraphics[width=0.7\linewidth]{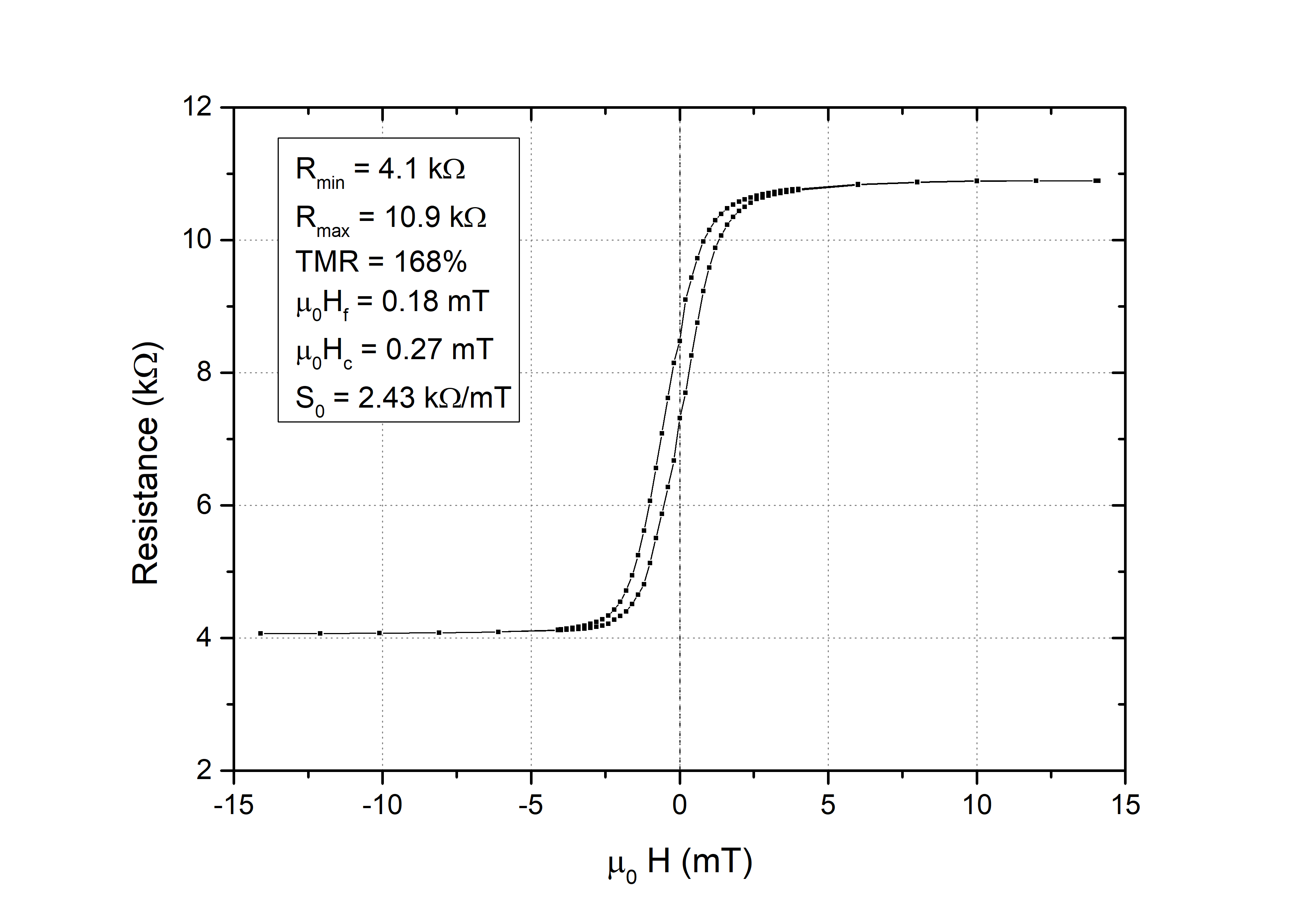}
\caption{Transfer curve of the TMR sensor used obtained through magnetotransport characterization, using a bias current of 100 $\mu m$.}
\end{figure}

As mock-up magnetic tracks, samples with micropatterned CoCrPt structures (denominated scales) were used, as shown in Figure 4. These microfabricated structures provide a clear advantage over commercial tracks as their accuracy is lower than $1 \mu m$ when compared to a minimum feature size of $80 \mu m$ for elastomer tracks \cite{bogen2019}. As such, CoCrPt scales provide controlled input parameters that can be used to validate the simulation model, although they cannot be used for large volume production. As such, a film with a nominal thickness of 2000 \AA of Co$_{66}$Cr$_{16}$Pt$_{18}$ was deposited on a glass substrate using a sputtering system (Alcatel SCM 450) and the patterns were then defined by optical lithography (Heidelberg DWL 2.0) and the excess material was removed through ion milling (Nordiko 3600). Using this process, eleven scales with 1000 $\mu m$ wide CoCrPt structures were obtained, with separations between 63 and 3000 $\mu m$.

\begin{figure}[h!]
\label{fig:4}
\centering\includegraphics[width=0.5\linewidth]{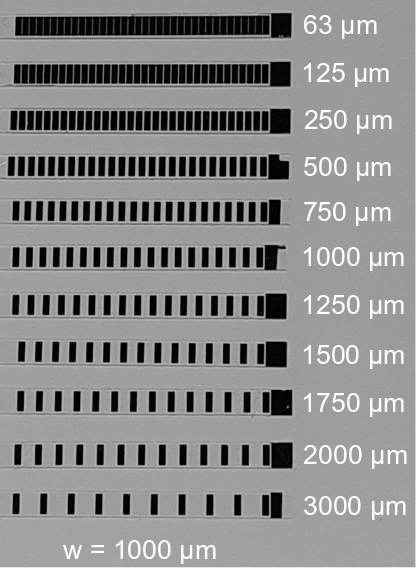}
\caption{Microfabricated CoCrPt scales. All structures have a width of 1000 $\mu m$, a length of 3000 $\mu m$ and their separation is indicated next to each scale in the figure.}
\end{figure}

\section{Simulation}
\label{S:3}

Considering the aforementioned PM properties, the magnetic field created by both of them placed in an antiparallel configuration and separated by 6 mm was simulated using finite elements modelling (FEM). This showed that the strength of the generated magnetic field was in the 71-72 mT range for reading distances between 100 and 1000 $\mu m$, which would then correspond to a magnetization of the CoCrPt structures in the 50-53 kA/m range, depending on the reading distance.

\begin{figure}[h!]
\label{fig:5}
\centering\includegraphics[width=\linewidth]{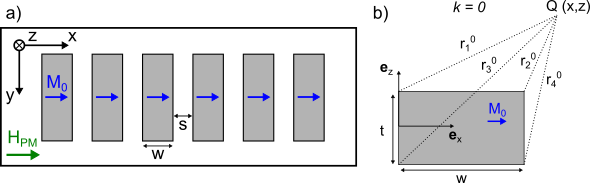}
\caption{a) Schematic representation of a scale with six CoCrPt structures (top view). b) Geometric parameters of a single structure when the magnetic stray field is calculated in a generic point Q (side view).}
\end{figure}

To simulate the magnetic stray field created by each of the CoCrPt structures, represented in Figure 5a, we firstly consider the equation for the magnetic stray field created by a generic magnetized volume \cite{bertram1994}:

\begin{equation}
\label{eq:main}
\mathbf{H}(\mathbf{r}) = - \frac{1}{4\pi} \int_V d^3\mathbf{r'} \left[\nabla\cdot\mathbf{M}(\mathbf{r'})\right] \frac{\mathbf{r} - \mathbf{r'}}{|\mathbf{r} - \mathbf{r'}|^3} + \frac{1}{4\pi} \int_S d^2\mathbf{r'} \left[\mathbf{M}(\mathbf{r'})\cdot \mathbf{e}_n\right] \frac{\mathbf{r} - \mathbf{r'}}{|\mathbf{r} - \mathbf{r'}|^3}
\end{equation}

$\mathbf{M}$ corresponds to the volumetric magnetization, $\mathbf{r}$ is the position vector of the point where the magnetic field is being calculated, $\mathbf{r'}$ is the position vector of the infinitesimal charge being considered and $\mathbf{e_n}$ is the outward surface normal at $\mathbf{r'}$. 

Considering that the magnetization of each structure is constant over the entire volume, then $\nabla\cdot\mathbf{M} = 0$, which removes the volumetric dependence. Furthermore, given the geometry of the system, it can be assumed that the properties of the structures along the y axis are constant. Therefore, a 2D approach can be followed to solve equation 1, which gives the following result for a CoCrPt scale with $N$ structures:

\begin{equation}
\label{eq:Hz}
H_z(x,z) = \frac{M_0}{2\pi} \sum^{N-1}_{k=0} \left(\log{\frac{r^k_4}{r^k_2}}-\log{\frac{r^k_3}{r^k_1}}\right)
\end{equation}

The geometric variables $r_i^k$ are represented in Figure 5b. This result is very similar to a previous one obtained by Merazzo et al. \cite{merazzo2018} However, in this case, the equations that describe the variables $r_i^k$ need to include bit separations different than the bit width. Therefore, they are represented as shown in the equation that follows, where $w$, $s$ and $t$ are the width, separation and thickness of the structures, respectively.

\begin{equation}
\label{eq:rk}
\left\{
\begin{array}{lr}
\mathbf{r}_1^k (x,z) = [x -kw - ks]\ \mathbf{e_x} + \left(z-\frac{t}{2}\right)\ \mathbf{e_z}\\
\mathbf{r}_2^k (x,z) = [x - (k+1)w - ks]\ \mathbf{e_x} + \left(z-\frac{t}{2}\right)\ \mathbf{e_z}\\
\mathbf{r}_3^k (x,z) = [x - kw - ks]\ \mathbf{e_x} + \left(z+\frac{t}{2}\right)\ \mathbf{e_z}\\
\mathbf{r}_4^k (x,z) = [x - (k+1)w - ks]\ \mathbf{e_x} + \left(z+\frac{t}{2}\right)\ \mathbf{e_z}\\
\end{array}
\right.
\end{equation}

This model will be used in the next section to compare with experimental results, both relatively to the trend of the curves and to their peak-to-peak amplitude. It is relevant to note that this model does not include magnetic phenomena such as dipolar interactions between structures in the same scale, which could be predominant for smaller separations.

\section{Results and discussion}
\label{S:4}

\subsection{Magnetic scans}
A 2D magnetic scan of seven different scales placed side-by-side is shown in Figure 6. Each scale can be clearly identified as well as the individual structures within a scale. Distortion of the magnetic field is visible in the boundary regions that connect two tracks, resultant from the interactions between structures from different scales. The analysis of this effect can be of particular interest for research involving absolute magnetic scales, although it is not the focus of the current study. 1D scans of isolated scales with different reading distances have also been obtained as well as equivalent simulated results. Table 1 shows the input parameters considered for the simulation. 

\begin{figure}[h!]
\label{fig:6}
\centering\includegraphics[width=0.8\linewidth]{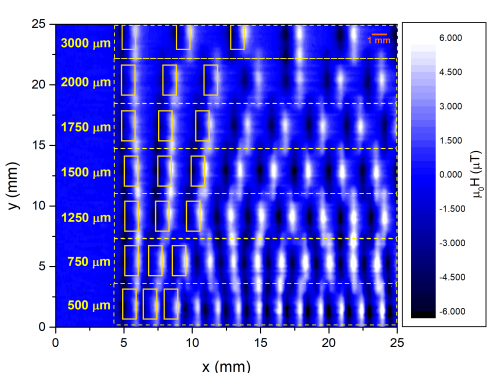}
\caption{2D magnetic scan of seven different CoCrPt scales obtained with a reading distance of 250 $\mu m$. The bit separation is written next to each scale and the structures have an area of 1 $\times$ 3 mm$^2$. The yellow dashed lines delimit each scale and the continuos yellow lines represent the location of each structure.}
\end{figure}

\begin{table}[h!]
\label{table:1}
\centering\begin{tabular}{lc}
\hline
Simulation parameter & Value                                 \\ \hline
Thickness            & 2202\ \AA \\
Structure width      & 1000 $\mu m$                          \\
Structure separation & 63 - 3000 $\mu m$                     \\
Magnetization        & 50 - 53 $kA/m$                          \\
Resolution           & 10 $\mu m$                           
\end{tabular}
\caption{Simulation parameters of the CoCrPt scales. The resolution corresponds to the difference in position between each simulated point.}
\end{table}

\begin{figure}[h!]
\label{fig:7}
\centering\includegraphics[width=0.8\linewidth]{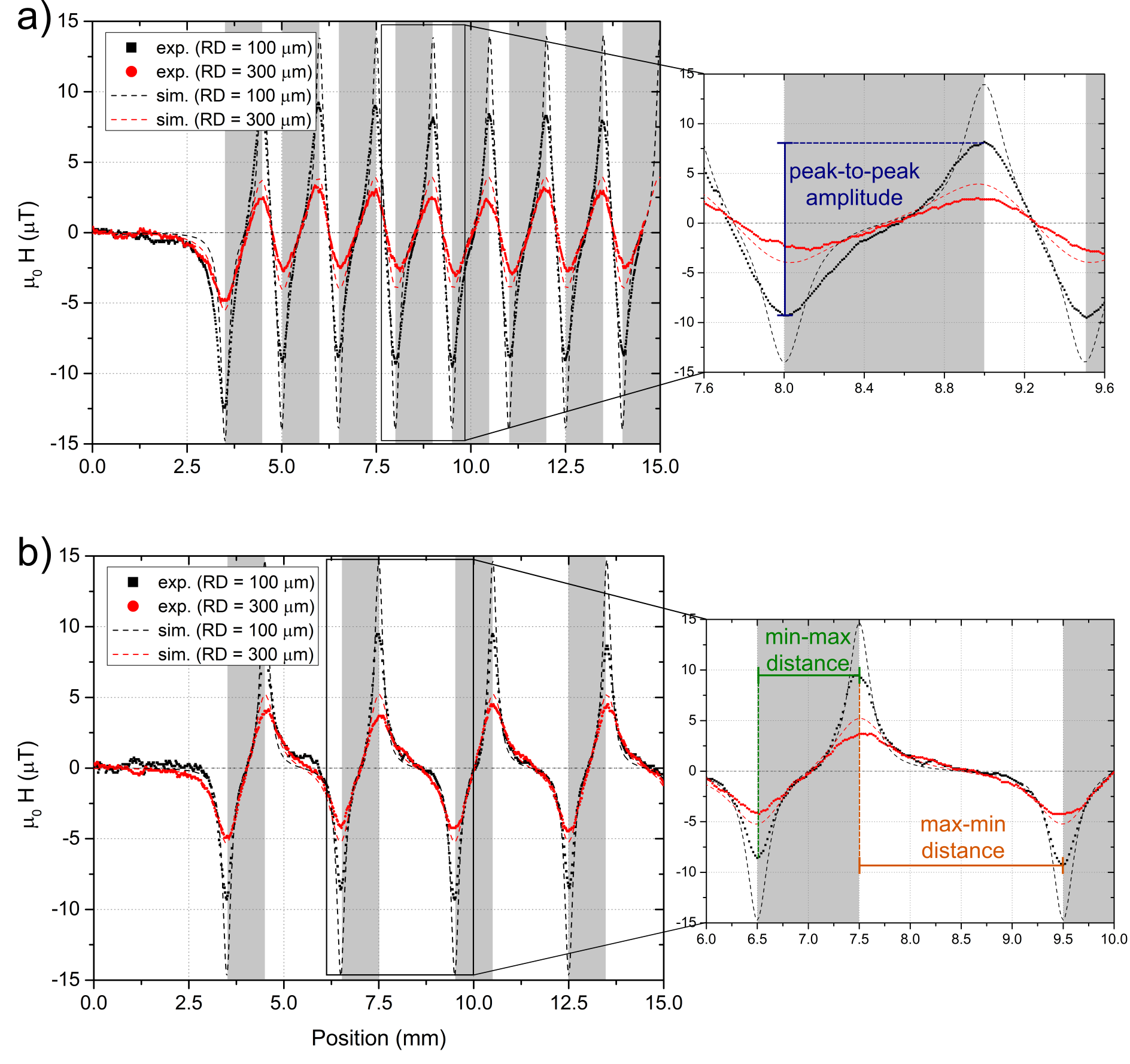}
\caption{Simulated and experimental results of magnetic scans of CoCrPt scales at two different reading distances (100 $\mu m$ and 300 $\mu m$). The grey areas represent the location of the structures. a) Structure width = 1000 $\mu m$, separation = 500 $\mu m$. b) Structure width = 1000 $\mu m$, separation = 2000 $\mu m$.}
\end{figure}
\pagebreak

Figure 7 contains scans of two representative scales with a structure width of 1000 $\mu m$, structure separations of 500 and 2000 $\mu m$ and a nominal thickness of 2000\ \AA, performed with reading distances of 100 and 300 $\mu m$. Simulated curves for these parameters are also included.

Qualitatively, it is possible to observe a match between the trends of both curves, with both signals having the same period, although the peak-to-peak amplitudes of the experimental signals are visibly smaller. This occurred for all the measured scales. In order to be able to quantitatively compare both results, it was necessary to determine both the position and the height of the peaks in each curve. With these results, it was possible to calculate both the average distance between consecutive extrema and the average peak-to-peak amplitude of each scan.

\subsection{Distance between consecutive extrema}

As represented in Figure 7, the ascending slopes of the peaks correspond to portions of the scale that contain CoCrPt structures and the descending ones correspond to regions without magnetic structures. As such, it is to be expected that the distance between a minimum and a maximum corresponds to the structure width and the distance between a maximum and a minimum corresponds to the separation. 

\begin{figure}[h!]
\label{fig:8}
\centering\includegraphics[width=0.7\linewidth]{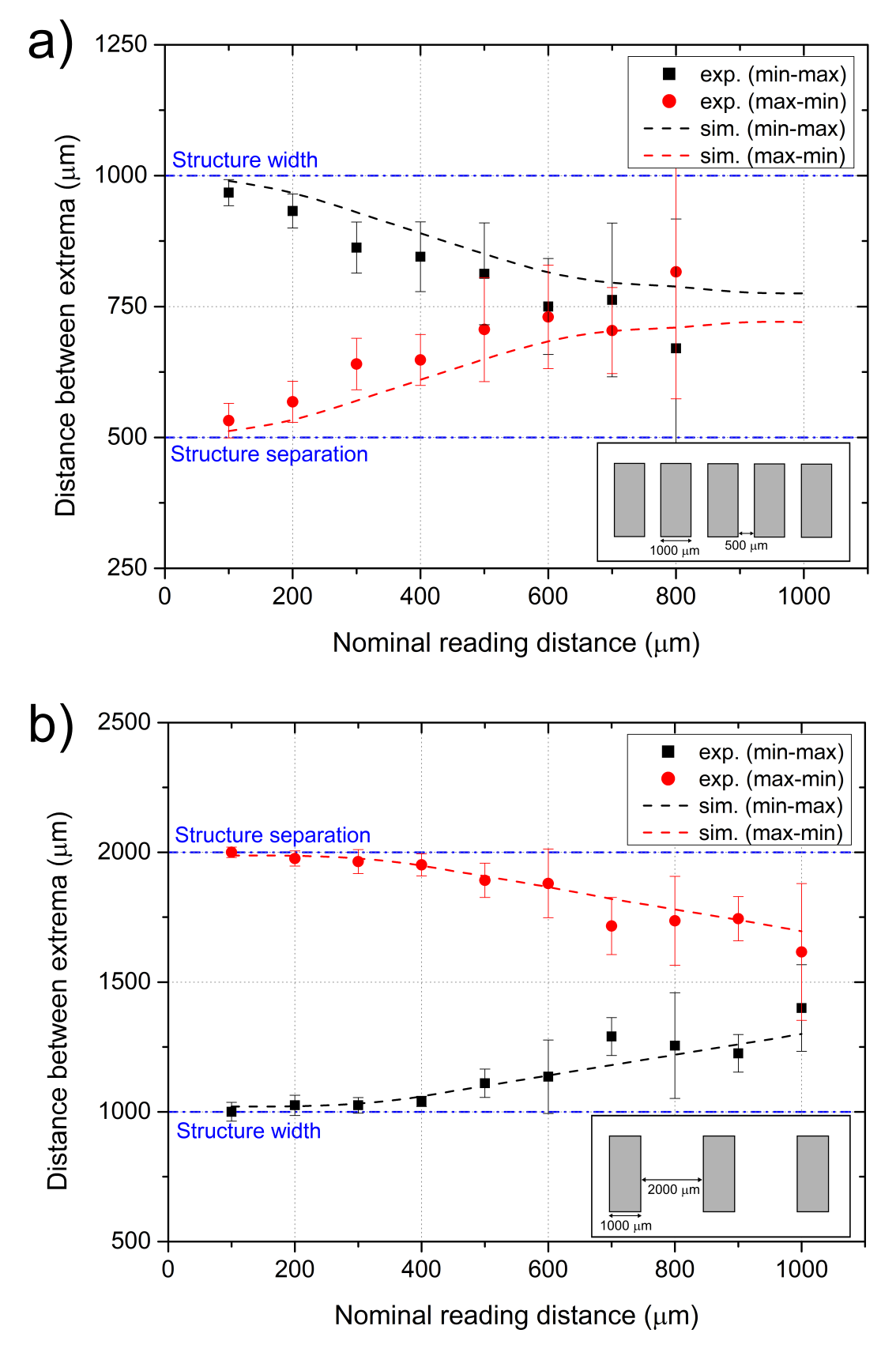}
\caption{Average distance between consecutive extrema (minimum to maximum and maximum to minimum) of each magnetic scan as a function of the reading distance. a) Structure width = 1000 $\mu m$, separation = 500 $\mu m$. b) Structure width = 1000 $\mu m$, separation = 2000 $\mu m$.}
\end{figure}

However, it was observed that this occurrence was only valid for measurements performed at a reading distance smaller than a threshold dependent on the separation. Figure 8 shows the distance between consecutive extrema (maximum to minimum and minimum to maximum) for two scales with separations 500 and 2000 $\mu m$ and reading distances between 100 and 1000 $\mu m$. For the scale with 500 $\mu m$ of separation, the distance between extrema never matches the expected values, and a convergence is observed. In the case of the scale with 2000 $\mu m$ of separation, these values match until a threshold reading distance of 400 $\mu m$, after which the values also start converging.  The theoretical model presented predicts these results, as represented by the dashed lines in the figure. Additionally, this behaviour was observed in all measured samples, except for the scale with the structure width equal to the separation. In this case, the distance between consecutive extrema was constant and equal to 1000 $\mu m$ when taking the measurement error into account.

This occurrence may prove to be a limitation for applications that depend on the location of each peak to determine accurate positions. As the spatial accuracy degrades above a specific reading distance, applications must guarantee that measurements occur under this value, which is defined by the feature sizes of the sample. 

\subsection{Peak-to-peak amplitude}

For the scales above, the peak-to-peak amplitude was calculated for the same interval of reading distances and the results are shown in Figure 9, along with a predicted curve from the theoretical model. This result is consistent for the remaining scales.

\begin{figure}[h!]
\label{fig:9}
\centering\includegraphics[width=0.7\linewidth]{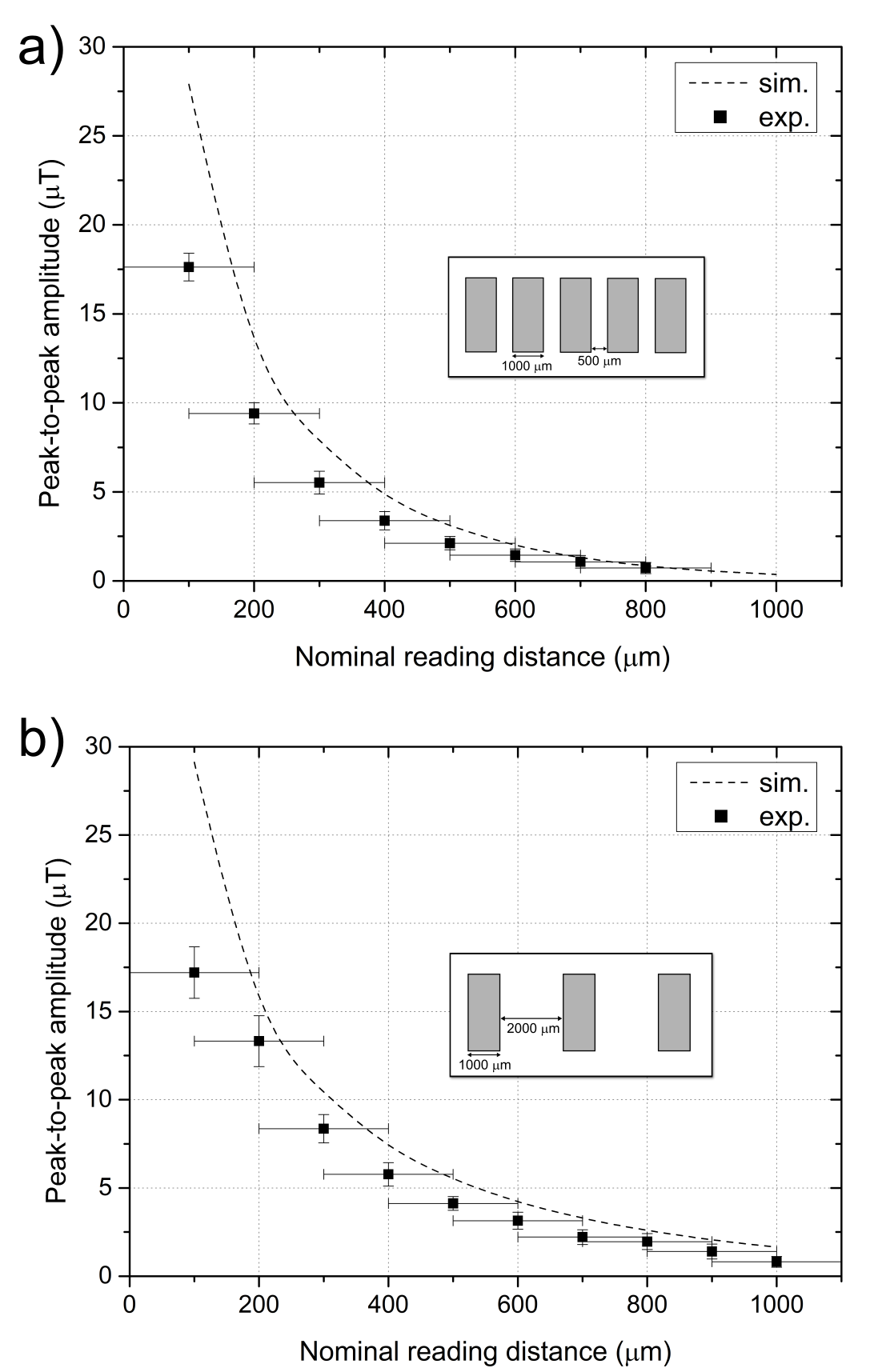}
\caption{Average peak-to-peak amplitude of each magnetic scan as a function of the reading distance. a) Structure width = 1000 $\mu m$, separation = 500 $\mu m$. b) Structure width = 1000 $\mu m$, separation = 2000 $\mu m$.}
\end{figure}

It can be observed that the experimental points and the theoretical curve are a partial match, when taking into account the error bars of the reading distance. This suggests that the model correctly describes the evolution of the peak-to-peak amplitude of the reading distance. However, the experimental points are always located under the theoretical curve. This may indicate that the experimental error of the reading distance is actually larger than 100 $\mu m$, leading to an overestimation of the reading distance in all measurements. This could be solved with an experimental setup with a better accuracy, ideally under $50 \mu m$, as the lowest reading distance considered in this study was 100 $\mu m$. On the other hand, it is also possible that other magnetic phenomena are present in the system and are not accounted for in the simulations, namely dipolar interactions. Furthermore, the model in use considers that the magnetization of each structure is constant, which has not been verified experimentally. The study of the dipolar interactions and the inclusion of magnetization variations along a structure would be necessary steps in the future to obtain a complete simulation model for this system.

\section{Conclusions}
\label{S:5}

For an incremental encoder system comprised of a CoCrPt scale and a reading head containing two permanent magnets and a TMR sensor, a theoretical model was developed. Scales with a nominal thickness of 2000\ \AA, with 1000 $\mu m$ wide structures and separations ranging from 63 to 3000 $\mu m$ were scanned with reading distances comprised in the 100-1000 $\mu m$ range. 

The trends of the experimental and simulated curves matched for all the scans performed, with the ascending slope of the peak corresponding to a portion of the track with a magnetic structure and the descending slope corresponding to one without. However, the peak-to-peak amplitudes of the signals were lower for the experimental results. This can mainly be due to inaccuracies when setting the reading distance in the experimental setup or dipolar couplings that are not accounted for in the proposed model.

For lower reading distances, the distances between consecutive extrema matched the structure width and separation. As such, we prove that a device utilizing differently sized structures can be used for positioning and similar applications, provided that the operational reading distance is lower than a threshold that depends on the minimum feature size. This is the first step towards being able to read more intricate incremental patterns than equally sized and spaced bits. The next stage would be to read tracks comprised of non-periodic and non-predetermined patterns or alphanumeric characters, which would create more complex signals to be interpreted.

Finally, although this work has been mainly focused on incremental tracks, the usage of CoCrPt scales can be extended to absolute tracks. Placing two or more of these scales side-by-side creates a system with well controlled dimensions and structure magnetization to simulate and test absolute measurement solutions.

\section*{Acknowledgements}

This work was supported by the Portuguese Foundation of Science and Technology (SFRH/BD/111538/2015), the National Infrastructure Roadmap (NORTE-01-0145-FEDER-22090), the European Regional Development Fund (LISBOA-01-0145-FEDER-031200) and the German Federal Ministry of Education and Research through the project GePos (02P16K000).



\bibliographystyle{model1-num-names}






\pagebreak

\textbf{Sofia Abrunhosa} was born in Lisbon, Portugal in 1994. In 2018, she received her MSc degree in Physics Engineering from Instituto Superior Técnico, University of Lisbon, Portugal. Her dissertation, entitled "Magnetoresistive sensors for industrial applications", was carried out at INESC Microsystems and Nanotechnologies in collaboration with BOGEN Electronic GmbH. She is now a research fellow at INESC-MN and is currently involved in the optimization process of magnetic tunnel junctions. Her research interests include thin film engineering, microfabrication of magnetic sensors and their implementation in automation, robotics and other industrial applications.\\

\textbf{Karla J. Merazzo}, a Costa Rican national, received her B.S. degree in Physics at the University of Costa Rica. There, she obtained teaching experience while working as an intern at Ad Astra Rocket Company in Liberia, Costa Rica, where she gained experience in plasma physics. She obtained a M.Sc and PhD degree at the Universidad Autónoma de Madrid, together with the Instituto de Ciencia de Materiales de Madrid (ICMM), CSIC, where she studied the properties of magnetic nanostructures such as nanowires and antidots arrays. She carried out postdoctoral studies at Spintec, Commissariat à l'Énergie Atomique (CEA), Grenoble, where she specialized in magnetic tunnel junctions (MTJ) for Spin Torque Nano-Oscillators. During her postdoc at INESC-MN in Lisbon, she worked on a different application of MTJs: magnetic sensors. All her work is experimental, including nanofabrication at clean room environment, electric-magnetic characterization and application of the devices. Currently, she is a consultant for SMEs, doing research, development and innovation of financing applications within the Horizon 2020 programs, where the core objective is to take fundamental devices or prototypes into the market.\\

\textbf{Tiago Costa} studied Physics Engineering in Instituto Superior Técnico, University of Lisbon, where he graduated in 2017. The last year of his studies was spent at INESC-MN, where he developed his thesis entitled “Advanced magnetoresistive sensors for industrial applications”, in which he participated actively in the development of an encoder system based on magnetoresistive technology in the scope of a collaboration with BOGEN Electronic GmbH. Since 2018 he has been working as a Research Engineer for this company, where he continues to develop encoding and measurement systems and ultimately push the use of state-of-the-art magnetic sensors into day-to-day applications.\\

\textbf{Oliver Sandig} joined BOGEN in 2017. He is the research manager and responsible for the optimization of the magnetic writing and reading process; this includes research into optimal head design, magnetic simulation and magnetic materials. Having focused on magnetism ever since his diploma thesis at Humboldt University zu Berlin, Oliver is a recognized specialist in the field. Prior to his work at Bogen, Oliver was a Ph.D. candidate at FU Berlin and Bessy II in Adlershof - Germany where he focused on the laser induced domain wall motion imaged with X-PEEM and studied the ultrafast demagnetisation of ultrathin magnetic films with TR-MOKE.\\

\textbf{Fernando Franco} was born in Alcobaça, Portugal in 1991. He received the B.S. and M.S. degrees in Physics Engineering from Instituto Superior Técnico, University of Lisbon, Portugal, in 2014 with a dissertation on non-destructive eddy current testing systems based on high sensitivity magnetoresistive sensors. He is currently pursuing the Ph.D. degree in Physics Engineering at the same university and a researcher at INESC Microsystems and Nanotechnologies with a project entitled “Development of a magnetic translator based on ultra-high resolution magnetoresistive sensors”. His research is focused on the development and optimization of highly sensitive magnetoresistive sensors for accurate magnetic encoders used in harsh industrial environments in collaboration with BOGEN Electronic GmbH.\\

\textbf{Susana Cardoso de Freitas} was born in Lisbon in 1973. She received the Ph.D. degree in Physics in 2002, from the Instituto Superior Técnico, Universidade de Lisboa. In 2002 she was a “Co-op Pre-Professional Engineer” at IBM, T.J.Watson Research Center (USA). In 2002 she became a researcher at INESC-MN and the co-leader of the Magnetics \& Spintronics group since 2006 with intensive collaborations with industrial partners for technology transfer related with sensors. She is an Associated Professor at the Physics Department (IST) since 2015, and is responsible for student coordination and educational activities related with nanoelectronics, microfabrication and spintronics. She has been invited for several international conferences and is co-author of over 290 publications (researcher ID: B-6199-2013). Her research interests include advanced thin films, microfabrication processes in large area wafers, and spintronic sensors for robotics, biomedical and industrial applications.

\end{document}